\documentclass[prd,aps,showpacs,nofootinbib,%preprint,
eqsecnum]{revtex4}
\usepackage{amsmath}
\usepackage{amssymb}
\usepackage{amsfonts}
\usepackage{graphicx,bm}
\usepackage{dcolumn}
\usepackage{color,amsxtra}
\usepackage{epsf}
\usepackage{enumerate}
\usepackage{hhline}
\usepackage{array}
\usepackage{tabularx}
\newcommand{\be}{\begin{equation}}
\newcommand{\ee}{\end{equation}}
\newcommand{\bea}{\begin{eqnarray}}
\newcommand{\eea}{\end{eqnarray}}
\newcommand{\beaa}{\begin{eqnarray*}}
\newcommand{\eeaa}{\end{eqnarray*}}

\newcommand{\nn}{\nonumber \\}
\newcommand{\e}{\mathrm{e}}

\def\be{\begin{equation}}
\def\ee{\end{equation}}
\def\bea{\begin{eqnarray}}
\def\eea{\end{eqnarray}}

\def\nn{\nonumber \\}
\def\e{\mathrm{e}}

\begin{document}

\tolerance=5000

\title{Reconstruction of Domain Wall Universe and Localization of Gravity}

\author{Masafumi Higuchi$^{1, }$\footnote{
E-mail address: mhiguchi@th.phys.nagoya-u.ac.jp}, 
Shin'ichi Nojiri$^{1, 2,}$\footnote{E-mail address:
nojiri@phys.nagoya-u.ac.jp}
}

\affiliation{
$^1$ Department of Physics, Nagoya University, Nagoya
464-8602, Japan \\
$^2$ Kobayashi-Maskawa Institute for the Origin of Particles and
the Universe, Nagoya University, Nagoya 464-8602, Japan}

\begin{abstract}

We construct a four-dimensional domain wall universe by using the Brans-Dicke type 
gravity with two scalar field. 
We give a formulation where for arbitrarily given warp factor and scale factor, 
we construct an action which reproduces both of the warp and scale factors 
as a solution of the Einstein equation and the field equations given by the action. 
This formulation could be called as reconstruction.    
We show that the model does not contain ghost with negative kinetic term and there 
occur the localization of gravity as in the Randall-Sundrum model. 
It should be noted that in the equation of the graviton, there appear extra terms 
related with the extra dimension, which might affect the tensor mode in the fluctuations 
in the universe. 

\end{abstract}


\maketitle

\section{Introduction \label{I}}

Many scenarios that our universe could be a brane in the higher dimensional 
space-time have been proposed 
\cite{Akama:1982jy,Randall:1999ee,Randall:1999vf,Dvali:2000hr,Deffayet:2000uy,Deffayet:2001pu}. 
Especially trace anomaly was used for the inflationary brane world models \cite{Nojiri:2000eb,Hawking:2000bb,Nojiri:2000gb}. 
The brane may be given by a limit where the thickness of the domain wall vanishes. 
In fact the domain world scenario, where we live on on the domain wall with finite thickness (thick brane), 
was proposed in \cite{Lukas:1998yy} before the proposition of the brane world scenario and 
bent domain wall \cite{Kaloper:1999sm} as well as 
dynamical domain wall \cite{Chamblin:1999ya} were also investigated. 
After that there were many activities in the domain wall or thick brane universe 
scenario \cite{DeWolfe:1999cp,Gremm:1999pj,Csaki:2000fc,Gremm:2000dj,Kobayashi:2001jd,Slatyer:2006un}. 
Recently in \cite{Toyozato:2012zh}, it has been proposed a domain wall model with two scalar fields and it has been 
shown that we can construct a model which generates space-time, 
where the scale factor of the domain wall universe, which could be the general FRW 
universe, and the warp factor are arbitrarily given. 
Such a formulation can be a generalization of the reconstruction of the domain wall \cite{Nojiri:2010wj}. 
A formulation, where only the warp factor of the domain wall can be arbitrary,  was 
proposed in the seminal paper \cite{DeWolfe:1999cp}. 
In the model \cite{Toyozato:2012zh}, the scalar field equations can be identified with the Bianchi identities: 
$\nabla^\mu \left( R_{\mu\nu} - \frac{1}{2}R g_{\mu\nu} \right)=0$ and therefore 
the equations can be satisfied automatically. 

A critical problem in the formulation of \cite{Toyozato:2012zh} is that there appears a ghost scalar field, 
whose kinetic term has non-canonical signature, in general although a model without ghost was proposed. 
In this paper, we show that we can construct a model which is ghost free and reproduces given scale and warp factors, 
by using the Brans-Dicke type model where a scalar field couples with the scalar curvature directly 
in the action. 
Furthermore we show that the graviton is localized on the domain wall and 
there appears massless graviton propagating on the domain wall in the 
model proposed in \cite{Toyozato:2012zh}. 
The massless graviton on the four dimensional brane world corresponds to the zero 
mode of the graviton in the five dimensional space-time. 
This tells that the graviton is localized on the domain wall even in the Brans-Dicke type model proposed in this paper. 
In the equation of the graviton, however, there appear extra terms 
related with the extra dimension, which may affect the tensor mode of the fluctuation in the universe. 

In the next section, we review on the formulation in \cite{Toyozato:2012zh}. 
In section \ref{III}, we construct the Brans-Dicke type model without ghost. 
In section \ref{IV}, we observe that the graviton is localized on the domain wall but 
in the equation of the graviton, there appear extra terms. 
The last section \ref{V} is devoted with the discussions. 

\section{Domain wall model with two scalar fields \label{II}}

In this section, we review the formulation of the domain wall model with two scalar fields 
based on \cite{Toyozato:2012zh}. 
The formulation is a generalization of the formulation in \cite{Bamba:2011nm}, where it was shown 
how we can construct models which admit the exact solutions describing the domain wall with given warp factor, 
and it was used a procedure for a scale factor in \cite{Capozziello:2005tf}. 

\subsection{Reconstruction of general FRW domain wall universe \label{IIIb}} 

We now investigate general domain wall, which can be regarded as a general FRW universe, and 
the. metric of the space-time in five dimensions is given by
\begin{align}
\label{metric}
ds^2 &= dw^2 + f \left( w, t \right) \left\{   \frac{dr^2}{1-kr^2} + r^2 d \theta^2 
+r^2 \sin^2 \theta d \phi^2  \right\} - \frac{e\left( w,t \right)^2}{f\left( w,t \right)} dt^2 
\nonumber \\
& \equiv \e^{\ln f \left( w,t \right)} \gamma_{mn} \left( x \right) dx^m dx^n 
+ h_{\alpha \beta} \left( y \right) dy^{\alpha} dy^{\beta} \, .
\end{align}
Here $m,n = 1,2,3$, $\alpha,\beta = 0,5$, and $y^0 = t$, $y^5= w$. 
By choosing
\be
\label{FRWchoice}
f \left( w,t \right) = L^2 \e^{u \left( w,t \right)} a \left( t \right)^2 \, ,\quad 
e \left( w, t \right) = L^2 \e^{u \left( w,t \right)} a \left( t \right) \, ,
\ee
we find that the general FRW universe, whose metric is 
\be
\label{FRWmetric0}
ds_\mathrm{FRW}^2 = - dt^2 + a \left( t \right)^2 \left\{   \frac{dr^2}{1-kr^2} + r^2 d \theta^2 
+r^2 \sin^2 \theta d \phi^2  \right\}\, ,
\ee
is embedded by the arbitrary warp factor $L^2 \e^{u \left( w,t \right)}$ in the five dimensional space-time. 

The following action with two scalar fields $\phi$ and $\chi$ was proposed in \cite{Toyozato:2012zh}: 
\be
\label{pc1}
S_{\phi\chi} = \int d^5 x \sqrt{-g} \left\{ \frac{R}{2\kappa^2} - \frac{1}{2} A (\phi,\chi) \partial_\mu \phi \partial^\mu \phi 
 - B (\phi,\chi) \partial_\mu \phi \partial^\mu \chi 
 - \frac{1}{2} C (\phi,\chi) \partial_\mu \chi \partial^\mu \chi - V (\phi,\chi)\right\}\, .
\ee
We can construct a model to realize the arbitrary metric written in the form of (\ref{metric}). 

For the model (\ref{pc1}), the energy-momentum tensor could be given by
\begin{align}
\label{pc2}
T^{\phi\chi}_{\mu\nu} =& g_{\mu\nu} \left\{ 
 - \frac{1}{2} A (\phi,\chi) \partial_\rho \phi \partial^\rho \phi 
 - B (\phi,\chi) \partial_\rho \phi \partial^\rho \chi 
 - \frac{1}{2} C (\phi,\chi) \partial_\rho \chi \partial^\rho \chi - V (\phi,\chi)\right\} \nn
& +  A (\phi,\chi) \partial_\mu \phi \partial_\nu \phi 
+ B (\phi,\chi) \left( \partial_\mu \phi \partial_\nu \chi + \partial_\nu \phi \partial_\mu \chi \right) 
+ C (\phi,\chi) \partial_\mu \chi \partial_\nu \chi \, .
\end{align}
On the other hand, the field equations read
\begin{align}
\label{pc3}
0 =& \frac{1}{2} A_\phi \partial_\mu \phi \partial^\mu \phi + A \nabla^\mu \partial_\mu \phi 
+ A_\chi \partial_\mu \phi \partial^\mu \chi 
+ \left( B_\chi - \frac{1}{2} C_\phi \right)\partial_\mu \chi \partial^\mu \chi  
+ B \nabla^\mu \partial_\mu \chi - V_\phi \, ,\\
\label{pc4}
0 =& \left( - \frac{1}{2} A_\chi + B_\phi \right) \partial_\mu \phi \partial^\mu \phi 
+ B \nabla^\mu \partial_\mu \phi 
+ \frac{1}{2} C_\chi \partial_\mu \chi \partial^\mu \chi 
+ C \nabla^\mu \partial_\mu \chi + C_\phi \partial_\mu \phi \partial^\mu \chi 
 - V_\chi\, .
\end{align}
Here $A_\phi=\partial A(\phi,\chi)/\partial \phi$, etc. 
We now choose $\phi=t$ and $\chi=w$. Then we find
\be
\label{pc4b}
T_0^{\ 0} = - \frac{f}{2e^2} A - \frac{1}{2} C - V\, ,\quad 
T_i^{\ j} = \delta_i^{\ j} \left( \frac{f}{2e^2} A - \frac{1}{2} C - V \right)\, ,\quad 
T_5^{\ 5} = \frac{f}{2e^2} A + \frac{1}{2} C - V\, ,\quad 
T_0^{\ 5} = B \, ,
\ee
and 
\begin{align}
\label{pc5}
0 =& - \frac{f}{2e^2} A_\phi + \frac{f}{e^2} \left( \frac{\dot e}{e} - \frac{2\dot f}{f} \right)A 
+ B_\chi + B \left( \frac{e'}{e} + \frac{f'}{f} \right) - \frac{1}{2} C_\phi - V_\phi \, ,\\
\label{pc9}
0 =& \frac{f}{2e^2} A_\chi - \frac{f}{e^2} B_\phi 
+ \frac{f}{e^2} \left( \frac{\dot e}{e} - \frac{2\dot f}{f} \right) B
+ \frac{1}{2} C_\chi + C \left( \frac{e'}{e} + \frac{f'}{f} \right) - V_\chi \, .
\end{align}
We may solve Eqs.~(\ref{pc4b}) with respect to $A$, $B$, $C$, $V$ and obtain 
\begin{align}
\label{pc7}
A =& \frac{e^2}{\kappa^2 f} \left( G_1^{\ 1} - G_0^{\ 0} \right) 
=  \frac{e^2}{\kappa^2 f} \left( G_2^{\ 2} - G_0^{\ 0} \right) 
=  \frac{e^2}{\kappa^2 f} \left( G_3^{\ 3} - G_0^{\ 0} \right) \nn
=& \frac{1}{\kappa^2} \left( - \frac{e^2 f''}{f^2} + \frac{e e''}{f} + \frac{2ke^2}{f^2} 
 - \frac{\ddot f}{f} + \frac{{\dot f}^2}{2 f^2} + \frac{\dot e \dot f}{ef} \right) \, , \nn
B =& \frac{1}{\kappa^2}G_0^{\ 5} = \frac{1}{\kappa^2} 
\left( - \frac{3 e' \dot f}{2 e^3} + \frac{3{\dot f}'}{2e^2} \right) \, , \nn
C =& \frac{1}{\kappa^2} \left( G_5^{\ 5} - G_1^{\ 1} \right)
= \frac{1}{\kappa^2} \left( G_5^{\ 5} - G_2^{\ 1} \right) 
= \frac{1}{\kappa^2} \left( G_5^{\ 5} - G_3^{\ 1} \right) \nn
=& \frac{1}{\kappa^2} \left( - \frac{f''}{2f} - \frac{e''}{e} - \frac{2k}{f} 
 - \frac{\ddot f}{2e^2} - \frac{{\dot f}^2}{2e^2 f} 
+ \frac{\dot e \dot f}{2e^3} + \frac{3 f' e'}{2fe} \right)\, , \nn
V =& \frac{1}{\kappa^2} \left( G_0^{\ 0} + G_5^{\ 5} \right) \nn
= & \frac{1}{\kappa^2} \left( - \frac{3f''}{4f} + \frac{3k}{f} 
+ \frac{3 {\dot f}^2}{4e^2 f} 
- \frac{3f'e'}{4fe} + \frac{3\ddot f}{4e^2} - \frac{3\dot e \dot f}{4e^3}
\right)\, .
\end{align}
Here $G_{\mu\nu}$ is the Einstein tensor. 
Therefore the equations (\ref{pc5}) and (\ref{pc9}) are nothing but the Bianchi identities:
\bea
\label{pc8}
&& - \frac{f}{2e^2} A_\phi + \frac{f}{e^2} \left( \frac{\dot e}{e} - \frac{2\dot f}{f} \right)A 
+ B_\chi + B \left( \frac{e'}{e} + \frac{f'}{f} \right) - \frac{1}{2} C_\phi - V_\phi 
= - \frac{e^2}{2f} \nabla^\mu G_\mu^{\ 0}\, ,\\
\label{pc10}
&& \frac{f}{2e^2} A_\chi - \frac{f}{e^2} B_\phi 
+ \frac{f}{e^2} \left( \frac{\dot e}{e} - \frac{2\dot f}{f} \right) B
+ \frac{1}{2} C_\chi + C \left( \frac{e'}{e} + \frac{f'}{f} \right) - V_\chi 
= \nabla^\mu G_\mu^{\ 5}\, .
\eea 
Therefore the equations (\ref{pc5}) and (\ref{pc9}) are automatically satisfied by choosing 
$A$, $B$, $C$, and $V$ by (\ref{pc7}). 

\subsection{An example of reconstructed model \label{SubIIb}}

In the action (\ref{pc1}), when one of the eigenvalue of the matrix 
$\left( \begin{array}{cc} A & B \\ B & C \end{array} \right)$ is negative, there appears 
a ghost, which conflicts with the quantum mechanics (see \cite{Nojiri:2013ru}, for example). 
In  \cite{Toyozato:2012zh}, some examples, where no ghost field appears, were given in 
case $k=0$, just for simplicity. 

As an example, we may assume $a(t) \propto t^{h_0}$ with a constant $h_0$. 
Then, the equations in (\ref{pc7}) give,
\begin{align}
\label{U1}
& \kappa^2 A = - \frac{\ddot U}{U} + \frac{3{\dot U}^2}{2 U^2} + \frac{h_0 \dot U}{t U} 
+ \frac{2h_0}{t^2}\, , \quad 
L^2 \kappa^2 B = - \frac{3U' \dot U}{2 U^3} + \frac{3 {\dot U}'}{2U^2} \, , \nn
& \kappa^2 C = - \frac{3 U''}{2U} + \frac{3 {U'}^2}{2 U^2} 
+ \frac{1}{L^2}\left( - \frac{\ddot U}{2 U^2} - \frac{5h_0 \dot U}{2 t U^2} 
 - \frac{3h_0^2 - h_0}{t^2U} \right)\, .
\end{align}
Here $U \left( w,t \right) \equiv\e^{u \left( w,t \right)}$. 
By further assuming $U \left( w,t \right) = W\left( w \right) T \left(t\right)$, Eq.~(\ref{U1}) 
can be rewritten as 
\begin{align}
\label{U2}
& \kappa^2 A = - \frac{\ddot T}{T} + \frac{3{\dot T}^2}{2 T^2} 
+ \frac{h_0 \dot T}{t T} + \frac{2h_0}{t^2}\, , \quad 
L^2 \kappa^2 B = 0 \, , \nn
& \kappa^2 C = - \frac{3 W''}{2W} + \frac{3 {W'}^2}{2 W^2} 
+ \frac{1}{L^2 WT^2}\left( - \frac{\ddot T}{2} - \frac{5h_0 \dot T}{2 t}
 - \frac{\left( 3h_0^2 - h_0\right) T}{t^2}\right)\, .
\end{align}
We now also assume $T \propto t^\beta$ and obtain 
\begin{align}
\label{U2B}
 - \frac{\ddot T}{2} - \frac{5h_0 \dot T}{2 t}
 - \frac{\left( 3h_0^2 - h_0\right) T}{t^2}
&\propto - \frac{1}{2} \left( \beta^2 - \left(1 - 5h_0 \right) \beta + 6 h_0^2 - 2 h_0 \right)  \nn
&= -\frac{1}{2} \left\{  \beta + \left(3 h_0 -1 \right) \right\} 
\left\{ \beta + 2 h_0 \right\} \, .
\end{align}
This tells that $T$ is given by
\be
\label{U3}
T(t) = T_1 t^{1 - 3 h_0} + T_2 t^{- 2h_0}\, ,
\ee
and we obtain
\be
\label{U5}
A = \frac{1}{\kappa^2} \left\{ \frac{3}{2}\left( \frac{\dot T(t)}{T(t)}
+ \frac{2h_0}{t} \right)^2  \right\}>0\, ,\quad 
\kappa^2 C = - \frac{3 W''}{2W} + \frac{3 {W'}^2}{2 W^2} 
= - \frac{3}{2} \left( \ln W \right)'' \, .
\ee
Then if $\ln W$ is convex, $C$ becomes positive. 
If we choose
\be
\label{U6}
W(w) = \e^{- \sqrt{1 + \frac{w^2}{w_0^2}}}\, ,
\ee
with a constant $w_0$, we find 
\be
\label{U7}
C = \frac{3}{2\kappa^2} \frac{\frac{1}{w_0^2}}{\left( 1 + \frac{w^2}{w_0^2} \right)^{\frac{3}{2}}} 
%\frac{3}{\kappa^2 w_0^2}
>0 \, .
\ee
Because both of $A$ and $C$ are positive and $B$ vanishes, any ghost does not appear 
in this model. 
Since $a(t) \propto t^{h_0}$,  the domain universe corresponds 
to the universe filled with the perfect fluid whose equation of state 
parameter $w$ is given by 
\be
\label{AAA3}
w = -1 + \frac{2}{3h_0}\, .
\ee
Since we now have $f \left(  w,t \right) = L^2 T \left( t \right) 
\e^{- \sqrt{1 + \frac{w^2}{w_0^2}}} a^2_0 t^{2h_0}$, 
$e \left(  w,t \right) = L^2 T \left( t \right) \e^{- \sqrt{1 + \frac{w^2}{w_0^2}}} a_0 t^{h_0}$ in 
(\ref{FRWchoice}). 
Then by using the last equation in (\ref{pc7}), we find the explicit form of the potential $V$:
\begin{align}
\label{AAAA1}
V  =&-\frac{3}{4\kappa^2}\left[\frac{W''}{W}+\frac{{W'}^2}{W^2}-\frac{1}{L^2WT}\left(\frac{\ddot{T}}{T}+\frac{5\dot{a}\dot{T}}{aT}
+\frac{2\ddot{a}}{a}+\frac{4\dot{a}^2}{a^2}\right)\right] \nn
=&-\frac{3}{4\kappa^2}\frac{1}{w_0^2}\left[2-\frac{2}{1+\frac{\chi^2}{w_0^2}}-\left(1+\frac{\chi^2}{w_0^2}\right)^{-\frac{3}{2}}\right] \, .
\end{align}
By replacing $w$ by $\chi$, we have obtained the explicit form 
of the potential $V \left( \chi  \right)$ in terms of the scalar field $\chi$. 
We should also note that we obtain a brane in the limit of $w_0\to 0$. 

\section{Brans-Dicke type model \label{III}}

As mentioned at the beginning of Subsection \ref{SubIIb}, the action (\ref{pc1}) includes ghost field in general. 
In this section, by using the Brans-Dicke type model, we consider a formulation where arbitrary FRW universe is realized 
without ghost. For the purpose, we use the example in Subsection \ref{SubIIb}, which does not include any ghost. 
When we consider the model, whose metric is given by scaling the metric in the model without ghost, the model 
does not include ghost, either. 

In this section we work by using the conformal time $\tau$, which is related with the cosmological time $t$ in 
(\ref{metric}) or (\ref{FRWmetric0}) by
\be
\label{BD1}
d\tau = \frac{dt}{a(t)}\, ,
\ee
and the FRW metric  (\ref{FRWmetric0}) can be rewritten as 
\be
\label{FRWmetric00}
ds_\mathrm{FRW}^2 = a\left(t\left(\tau\right) \right)^2 \left[ - d\tau^2 + \left\{  dr^2 + r^2 d \theta^2 
+r^2 \sin^2 \theta d \phi^2  \right\}\right]\, ,
\ee
In this section, we only consider $k=0$ case, for simplicity, again. 
In case of Subsection \ref{SubIIb}, $a(t) = a_0 t^{h_0}$ with constants $a_0$ and $h_0$ 
and we find the explicit relation between $\tau$ and $t$, as follows
\be
\label{BD2}
\tau = \frac{t^{1 - h_0}}{\left( 1 - h_0 \right) a_0}\, .
\ee
Then if we define a new scalar field $\varphi$ by 
\be
\label{BD3}
\varphi =\varphi(t) \equiv  \frac{\phi^{1 - h_0}}{\left( 1 - h_0 \right) a_0}\, ,
\ee
we find 
\be
\label{BD4}
\varphi=\tau\, ,\quad 
a(t) = \tilde a (\tau) = 
a_0 \left( \left( 1-h_0 \right)a_0\tau \right)^{\frac{h_0}{1-h_0}}\, .
\ee
Then arbitrary warp factor $ a\left(t\left(\tau\right) \right) = A(\tau)$ can be realized by multiplying a function with the 
metric $g_{\mu\nu}$ by $\frac{A\left(\tau\right)^2}{\tilde a\left( t\left(\tau\right)\right)^2}=\e^{2\Theta(t)}$ or 
\be
\label{BD5}
g_{\mu\nu} \to \e^{-2\Theta(\phi)} g_{\mu\nu} \, .
\ee
Then since
\be
\label{BD6}
R\to \e^{2\Theta(\phi)}  \left( R + 8 \Box \Theta (\phi) -12 \partial_\mu \Theta (\phi) \partial^\mu \Theta (\phi) \right) \, ,
\ee
in five dimensions, the action (\ref{pc1}) can be rewritten as
\begin{align}
\label{pc1BD}
S_{\phi\chi} =& \int d^5 x \sqrt{-g} \left\{ \frac{\e^{-3\Theta(\phi)} R}{2\kappa^2} - \frac{1}{2} \e^{-3\Theta(\phi)} \left( A (\phi,\chi) 
 - \frac{12}{\kappa^2} \Theta'(\phi)^2 \right) \partial_\mu \phi \partial^\mu \phi 
 - \e^{-3\Theta(\phi)} B (\phi,\chi) \partial_\mu \phi \partial^\mu \chi \right. \nn
& \left. - \frac{1}{2} \e^{-3\Theta(\phi)} C (\phi,\chi) \partial_\mu \chi \partial^\mu \chi - \e^{-5\Theta(\phi)} V (\phi,\chi) \right\} \nn
 =&\frac{1}{2\kappa^2}  \int d^5x \sqrt{-g}
\left\{\left(\frac{a_0\phi^{h_0}}{A(\phi)}\right)^3R-3\left(\frac{a_0\phi^{h_0}}{A(\phi)}\right)^3\left[\frac{1}{2}
\left(\frac{\dot{T}(\phi)}{T(\phi)}+\frac{2h_0}{\phi}\right)^2-4
\left( \frac{A'(\phi)}{A(\phi)} - \frac{h_0}{\phi} \right)^2 
\right]\partial_\mu\phi\partial^\mu\phi \right. \nn
& \left. -\frac{3}{2w_0^2}\left(1+\frac{\chi^2}{w_0^2}\right)^{-\frac{3}{2}}\left(\frac{a_0\phi^{h_0}}{A(\phi)}\right)^3
\partial_\mu\chi\partial^\mu\chi+\frac{3}{2w_0^2}\left(\frac{a_0\phi^{h_0}}{A(\phi)}\right)^5
\left[2-\frac{2}{1+\frac{\chi^2}{w_0^2}}-\left(1+\frac{\chi^2}{w_0^2}\right)^{-\frac{3}{2}}\right]\right\}\, .
\end{align}
This action can be regarded as the action in the Jordan frame action. 
Let assume that the particles in the standard model of the elementary particle physics couple with the metric in the Jordan frame and 
do not couple with the scalar field $\phi$ nor $\chi$.   
Then if we start with the action (\ref{pc1BD}), we can realized any history of the expansion of the universe given by 
 $ a\left(t\left(\tau\right) \right) = A(\tau)$. 

\section{Localization of graviton \label{IV}}

In the second Randall-Sundrum model \cite{Randall:1999vf}, the massless graviton is localized on the brane. 
The massless graviton corresponds to the zero mode of the graviton in five dimensions. 
In \cite{Toyozato:2012zh}, it has been shown that the graviton can be localized on the domain wall, which corresponds 
to the flat, de Sitter, or anti-de Sitter space-time. 
In this section, we show that the localization of the graviton could occur in the model (\ref{pc1}), which tells that the localization occurs 
even in the Brans-Dicke type model (\ref{pc1BD}), since we do not change the warp factor in the scale transformation (\ref{BD5}). 
An interesting point is that there appears a correction proportional to the derivative of the warp factor with respect the time in the 
equation of the graviton. This correction may affect the tensor mode of the flucutuation in the universe. 
Before going to the problem of the localization, we need to find the equation of the graviton in the FRW universe in four dimensions. 
In order to find the equation, we consider the model where a scalar field coupled with gravity. 
After that we compared the equation for the zero mode of the graviton in five dimensions with the equation of the graviton in 
four dimensions. 

\subsection{Equation of  graviton in four dimensional FRW universe}

We now consider the following perturbation
\be
\label{H1}
g_{\mu\nu} \to g_{\mu\nu} + h_{\mu\nu}\, .
\ee
Then we obtain
\begin{align}
\label{H2}
\delta R_{\mu\nu} = & \frac{1}{2} \left[ \nabla_\mu \nabla^\rho h_{\nu\rho} 
+ \nabla_\nu \nabla^\rho h_{\mu\rho} - \nabla^2 h_{\mu\nu} 
 - \nabla_\mu \nabla_\nu \left( g^{\rho\lambda} h_{\rho\lambda} \right) 
 - 2 R^{\lambda\ \rho}_{\ \nu\ \mu} h_{\lambda\rho} 
+ R^\rho_{\ \mu} h_{\mu\nu} + R^\rho_{\ \nu} h_{\rho\mu} \right] \, ,\\
\label{H3}
\delta R = & - h_{\mu\nu} R^{\mu\nu} + \nabla^\mu \nabla^\nu h_{\mu\nu} 
 - \nabla^2 \left( g^{\mu\nu} h_{\mu\nu}\right)\, .
\end{align}
By imposing the gauge condition
\be
\label{H4}
\nabla^\mu h_{\mu\nu} = g^{\mu\nu} h_{\mu\nu} = 0\, ,
\ee
the Einstein equation $R_{\mu\nu} - \frac{1}{2} g_{\mu\nu} R = \kappa^2 T_{\mu\nu}$ 
has the following form:
\be
\label{H6}
\frac{1}{2} \left[ - \nabla^2 h_{\mu\nu} 
 - 2 R^{\lambda\ \rho}_{\ \nu\ \mu} h_{\lambda\rho} 
+ R^\rho_{\ \mu} h_{\mu\nu} + R^\rho_{\ \nu} h_{\rho\mu}
 - h_{\mu\nu} R + g_{\mu\nu} R^{\rho\lambda} h_{\rho\lambda} \right] 
= \kappa^2 \delta T_{\mu\nu}\, .
\ee
We may consider scalar field theory in \cite{Nojiri:2005pu}, whose action is given by
\be
\label{H7}
S_\phi = \int d^4 x \sqrt{-g} \mathcal{L}_\phi\, , \quad 
\mathcal{L}_\phi =  - \frac{1}{2} \omega(\phi) \partial_\mu \phi 
\partial^\mu \phi - V(\phi) \, .
\ee
Then we obtain
\be
\label{H8}
T_{\mu\nu} = - \omega(\phi) \partial_\mu \phi \partial_\nu \phi 
+ g_{\mu\nu} \mathcal{L}_\phi\, ,
\ee
and therefore 
\be
\label{H9}
\delta T_{\mu\nu} = h_{\mu\nu} \mathcal{L}_\phi 
+ \frac{1}{2} g_{\mu\nu} \omega(\phi) \partial^\rho \phi \partial^\lambda \phi h_{\rho\lambda}\, .
\ee 

Because we are now interested in the graviton, we may assume $h_{\mu\nu}=0$ except 
the components with $\mu,\nu=1,2,3$. 
In the FRW universe (\ref{FRWmetric0}) with $k=0$, we may assume $\phi=t$ in (\ref{H9}), 
Then by using the FRW equation
\be
\label{LB1}
\frac{3}{\kappa^2} \frac{{\dot a}^2}{a^2} = \frac{\omega}{2} + V\, ,\quad 
\frac{1}{\kappa^2} \left( 2\frac{\ddot{a}}{a} + \frac{\dot{a}^2}{a^2}+\frac{k}{a^2}\right)
= \frac{\omega}{2} - V\, ,
\ee
we find 
\be
\label{LB2}
\omega = \frac{1}{\kappa^2} \left( 2\frac{\ddot{a}}{a} + 4 \frac{\dot{a}^2}{a^2}+\frac{k}{a^2}\right), 
V = -\frac{1}{\kappa^2} \left( \frac{\ddot{a}}{a} + \frac{\dot{a}^2}{a^2}+\frac{k}{a^2}\right)\, .
\ee
By using (\ref{H6}), (\ref{H9}), (\ref{LB2}), and $\phi=t$, we find the equation of graviton:
\be
\label{LB3}
0=\left(2\frac{\ddot{a}}{a}+\frac{\dot{a}}{a}\partial_t-\partial_t^2+\frac{\bigtriangleup}{a^2}\right)h_{ij}\, .
\ee

\subsection{Localization of the graviton on the domain wall}

In this subsection, we show that the graviton can be localized on the domain wall or there exists a zero mode solution in 
the equation for the graviton in five dimensions. 
Explicit formulas for the connection and curvatures are given in Appendix. 

By using (\ref{pc4b}) and (\ref{pc5}) or
\begin{align}
\label{LB4}
\kappa^2 A =& -\ddot{u}+\frac{1}{2}\dot{u}^2 -2\frac{\ddot{a}}{a}+2\frac{\dot{a}^2}{a^2}+\frac{\dot{a}\dot{u}}{a}+2\frac{k}{a^2} \, ,\nn
\kappa^2 B=&-\frac{3}{2}\dot{u}^\prime \, ,\nn
\kappa^2 C=&-\frac{3}{2}u^{\prime\prime}-\frac{1}{2}L^{-2} \e^{-u}\left(\ddot{u}+\dot{u}^2+2\frac{\ddot{a}}{a}
+5\frac{\dot{a}\dot{u}}{a}+4\frac{\dot{a}^2}{a^2}+4\frac{k}{a^2}\right) \, ,\nn
\kappa^2 V=&-\frac{3}{4}\left[u^{\prime\prime}+2u^{\prime2}-L^{-2} \e^{-u}\left(\ddot{u}
+\dot{u}^2+2\frac{\ddot{a}}{a}+4\frac{\dot{a}^2}{a^2}+5\frac{\dot{a}\dot{u}}{a}+4\frac{k}{a^2}\right)\right] \, ,
\end{align}
we find the equation for graviton in five dimensions:
\be
\label{LB5}
0= \left[\partial^2_w-u^{\prime\prime}-u^{\prime2}+L^{-2} \e^{-u}\left(\ddot{u}
+\frac{\dot{a}\dot{u}}{a}+\dot{u}\partial_t
+2\frac{\ddot{a}}{a}+\frac{\dot{a}}{a}\partial_t-\partial_t^2+\frac{\bigtriangleup}{a^2}\right)\right]h_{ij} \, .
\ee
By assuming $h_{ij}(w,x)= \e^{u(w,t)}\hat{h}_{ij}(x)$, we find the following equation: 
\be
\label{LB6}
0 = \left(2\frac{\dot{a}\dot{u}}{a}-\dot{u}\partial_t+2\frac{\ddot{a}}{a}+\frac{\dot{a}}{a}\partial_t-\partial_t^2
+\frac{\bigtriangleup}{a^2}\right)\hat{h}_{ij}\, .
\ee
Then if $u$ goes to minus infinity sufficiently rapidly for large $|w|$, $h_{ij}(w,x)$ is normalized in the direction of $w$ 
and therefore there occurs the localization of graviton. 

We should also note that if $\dot{u}\left(2\frac{\dot{a}}{a}-\partial_t\right)\hat{h}_{ij}=0$, the above expression (\ref{LB6}) coincides with 
the equation for the graviton in (\ref{LB3}). Especially when the warp factor does not depend on time, that is, $\dot u=0$, two 
expressions coincide with each other. 
Conversely if $\dot u \neq 0$, there could appear some corrections proportional to $\dot u$ when we consider the tensor 
perturbation. 

\section{Discussions \label{V}}

By using the Brans-Dicke type gravity with two scalar field, we constructed 
a four-dimensional domain wall universe. 
In the formulation, when the warp factor and scale factor are arbitrarily given, 
we can construct an action which reproduces both of the warp and scale factors 
as an exact solution of the Einstein equation and the field equations given by the action. 
The obtained model does not contain ghost with negative kinetic term and there 
occur the localization of gravity as in the Randall-Sundrum model. 
In the equation of the graviton, there appear extra terms 
related with the extra dimension, which is proportional to the derivative of the warp factor 
with respect to the cosmological time $t$. 
This extra terms might affect the tensor mode in the fluctuations 
in the universe and might be found by the observations of the CMB and/or 
the structure formation of the universe. 

The remaining problem is to investigate the stability of the domain wall solution, which 
requires the time-dependent perturbation 
from the solution. The existence of the massless graviton may tell 
that the model could be stable under the perturbation 
of the tensor mode in the metric. We need, of course, to include the 
perturbation of the scalar mode including the scalar fields $\phi$ and $\chi$, 
in order to verify the stability, which could be a future work. 

\section*{Acknowledgments.}

We are grateful to S.~D.~Odintsov for the discussion when he stayed in 
Nagoya University as a JSPS fellow. 
S.N. is supported by the JSPS Grant-in-Aid for Scientific Research (S) \# 22224003
and (C) \# 23540296.

\appendix

\subsection{Explicit expressions of connections and curvatures in five dimensions}

In this appendix, we give explicit expressions of connections and curvatures in five dimensional space-time, whose 
metric is given by
\begin{align}
\label{A1}
 g_{AB}=\begin{pmatrix}
 	-L^2 \e^{u(w,t)}&&&&\\
	 & L^2 \e^{u(w,t)}\frac{a(t)^2}{1-kr^2} & & & \\
	 & & L^2 \e^{u(w,t)}a(t)^2r^2 & & \\
	 & & & L^2 \e^{u(w,t)}a(t)^2r^2\sin^2\theta & \\
	 & & & & 1 
	\end{pmatrix}\, .
\end{align}\\
Then the connections are given by
\begin{align}
\label{A2}
&\Gamma^t_{tt}=\frac{1}{2}\dot{u}\, ,\quad
\Gamma^w_{tt}=\frac{1}{2}L^2e^uu^\prime\, , \quad
\Gamma^r_{rt}=\Gamma^\theta_{\theta t}=\Gamma^\phi_{\phi t}=\frac{\dot{a}}{a}+\frac{1}{2}\dot{u} \, , \quad
\Gamma^t_{tw}=\Gamma^r_{rw}=\Gamma^\theta_{\theta w}=\Gamma^\phi_{\phi w}=\frac{1}{2}u^\prime\, , \nn
&\Gamma^t_{ij}=L^{-2} \e^{-u}\left(\frac{\dot{a}}{a}+\frac{1}{2}\dot{u}\right)g_{ij}\, , \quad
\Gamma^r_{rr}=\frac{kr}{1-kr^2}\, , \quad
\Gamma^w_{ij}=-\frac{1}{2}u^\prime g_{ij}\, , \quad
\Gamma^\theta_{\theta r}=\Gamma^\phi_{\phi r}=\frac{1}{r}\, , \nn
&\Gamma^r_{\theta\theta}=-r(1-kr^2)\, , \quad
\Gamma^\phi_{\phi\theta}=\cot\theta\, , \quad
\Gamma^r_{\phi\phi}=-r(1-kr^2)\sin^2\theta\, ,\quad
\Gamma^\theta_{\phi\phi}=-\cos\theta\sin\theta \, .
\end{align}
The Ricci curvatures have the following forms:
\begin{align}
R_{tt}&=\left[-\frac{1}{2}u^{\prime\prime}-u^{\prime2}+\frac{3}{2}L^{-2} \e^{-u}\left(\ddot{u}+\frac{\dot{a}\dot{u}}{a}
+2\frac{\ddot{a}}{a}\right)\right]g_{tt}\, ,\nn
R_{ij}&=\left[-\frac{1}{2}u^{\prime\prime}-u^{\prime2}+\frac{1}{2}L^{-2} \e^{-u}\left(\ddot{u}+5\frac{\dot{a}\dot{u}}{a}
+2\frac{\ddot{a}}{a}+4\frac{\dot{a}^2}{a^2}+\dot{u}^2+4\frac{k}{a^2}\right)\right]g_{ij} \, ,\nn
R_{ww}&=-2u^{\prime\prime}-u^{\prime2} \, \nn
R_{tw}&=-\frac{3}{2}\dot{u}^\prime \, .
\end{align}
The scalar curvatures is
\begin{align}
\label{A3}
 R=-4u^{\prime\prime}-5u^{\prime2}+3L^{-2} \e^{-u}\left(\ddot{u}+\frac{1}{2}\dot{u}^2
+3\frac{\dot{a}\dot{u}}{a}+2\frac{\ddot{a}}{a}+2\frac{\dot{a}^2}{a^2}+2\frac{k}{a^2}\right)\, .
\end{align}

\end{document}